\documentstyle[aps,multicol]{revtex}
\begin{document}

\draft

\title{\bf RELATION BETWEEN ANDERSON- AND INVARIANT IMBEDDING MODELS FOR TRANSPORT IN FINITE CHAINS AND STUDY OF PHASE- AND DELAY TIME DISTRIBUTIONS FOR STRONG
DISORDER}
\author{J.Heinrichs}
\address{Institut de physique, B5, Universit\'{e} de Li\`{e}ge, Sart Tilman, B-4000 Li\`{e}ge, Belgium}
\maketitle
\date{\today}
\maketitle

\begin{abstract}
\thispagestyle{empty}
\noindent
The invariant imbedding evolution equations for the amplitude reflection and transmission coefficients of a disordered 1D chain are shown to follow from the
continuum limit, for weak disorder, of recursion relations between reflection (transmission) coefficients of Anderson chains of $N$ and $N-1$ atoms
embedded in infinite non-disordered chains.  We show that various available analytical and numerical results for reflection phase distributions in the Anderson-
and invariant imbedding models are in good qualitative agreement, as expected from the above equivalence of these models.  We also discuss the distribution of
reflection phases and of reflection delay times for an Anderson model with a simple distribution of on-site potentials characterizing extreme strong disorder.

\pacs{72.10.-d, 72.70.+m, 73.23.-b \\e-mail: J.Heinrichs@ulg.ac.be}

\end{abstract}

\pagestyle{plain}
\setcounter{page}{1}
\begin{multicols}{2} 
\section{INTRODUCTION}
In recent years a number of studies of phase-\cite{1,2,3,4,5} and delay time distributions\cite{4,6} for waves reflected by a finite 1D Anderson chain have been
published in parallel with other studies of phase-\cite{7,8,9,10,11,12} and delay time distributions\cite{9,13,14,15} for continuous disordered chains. In
particular, an important result which has emerged from the latest studies of delay-time distributions\cite{4,14,15} for different 1D systems, is the universality
of their asymptotic form,

\begin{equation}\label{eq1}
P_\tau (\tau,L\rightarrow\infty)={L_c\over k\tau^2}e^{-{L_c\over k\tau}},\;\tau\rightarrow\infty\quad ,
\end{equation}

\noindent
which was first derived in Refs \cite{9} and \cite{13} (see, in particular, the footnote~14 in Ref \cite{9} and noting that the continuum localization length
is actually twice the value listed in\cite{9} and \cite{10} (see Sect.II.C below)).  Here $\tau$ is the time delay, $L_c$ the localization length and $k$ the
wavenumber of a reflected wave.  Furthermore, the fact that the equation (\ref{eq1}), which coincides with the distribution predicted by random matrix theory
(RMT)\cite{16}, has been reobtained by Comtet and Texier\cite{14} for two~different kinds of disorder in the continuum case and by Ossipov {\it et al.}\cite{4}
for the Anderson model, strongly supports the conclusion that the universality of RMT also extends to strictly 1D~random systems where Anderson localization
dominates.

On the other hand, the earlier continuous space treatments\cite{8,9,10,11} of phase- and delay time distributions are based on an invariant imbedding
description\cite{17,18} of reflection and transmission properties of a 1D~disordered system.  We recall that the invariant imbedding method, developed originally
by S.~Chandrasekhar, leads to stochastic differential equations describing the evolution of the amplitude reflection (transmission) coefficient as a function of
length of a disordered sample, viewed as a macroscopic continuum\cite{17}.  The relationship between this method and corresponding studies for tight-binding
chains described by the Anderson model has remained somewhat elusive.

In view of the extensive use which has been made of both models of 1D~chains, it is useful to examine their possible connection in more detail.  In part~\ref{prta}
of Section~\ref{sec2} we present an exact derivation of the invariant imbedding evolution equations from recursion relations between amplitude reflection
(transmission) coefficients of tight-binding Anderson chains of $N$ and $N-1$ atoms\cite{3}, respectively, for weak disorder.

Next, in part~\ref{prtb}, we compare available analytical and/or numerical results for reflection phase distributions for the Anderson-\cite{2,4} and invariant
imbedding\cite{8,9,10,12} models of disordered chains, in order to see to what extent they reflect the relationship between these models established in
\ref{prta}.  In a similar spirit we briefly discuss in part~\ref{prtc} the effect of discreteness of the Anderson model on the statistical properties of the
reflection coefficient in a precise way, for weak disorder.

In the final section~\ref{sec3} we discuss the reflection phase- and delay-time distributions ana\-lytically for a special case of a strongly disordered
one-dimensional Anderson model.  Our specific assumption is that the random site energies $\varepsilon_i$ follow a doubly peaked distribution composed of narrow
peaks centered at large positive and negative values $\varepsilon_i\equiv\pm c$.  This distribution corresponds to a simple tractable case of extreme strong
disorder for a tight-binding chain.  The phase- and delay-time distributions obtained in this case differ qualitatively from results for the more familiar strong
disorder case with a wide symmetric distribution of site energies centered at $\varepsilon_i=0$\cite{4}.

\newpage
\section{TRANSPORT IN FINITE CHAINS \\ FOR WEAK DISORDER}\label{sec2}
\subsection{Derivation of invariant imbedding from Anderson model description}\label{prta}

Here we first recall the derivation of Barnes and Luck\cite{3} of a recursion relation for the amplitude reflection coefficient of a disordered chain of $N$ atoms
described by the Anderson model, in terms of the reflection coefficient of a chain with one less disordered atom.  We then extend this discussion for obtaining a
similar recursion relation for amplitude transmission coefficients of disordered chains.  These Anderson model recursion relations are the starting point for
deriving the invariant imbedding stochastic equation for weak disorder.  We also apply them in the study of reflection phase- and time-delay distributions in the
special case of strong disorder considered in Section~\ref{sec3}.

The Anderson model for a chain of $N$ disordered atomic sites (of spacing $a=1$), \linebreak $1\leq m\leq N$, embedded in an infinite non-disordered chain is
defined by the tight-binding Schr\"{o}dinger equation

\begin{equation}\label{eq2}
\varphi_{n+1}+\varphi_{n-1}+\varepsilon_n\varphi_n=E\varphi_n\quad ,
\end{equation}

\noindent
where the $\varphi_m$ denote the wavefunction amplitudes at sites $m$ and the sites energies $\varepsilon_m$ are random variables associated with the disordered
sites $1\leq m\leq N$.  Finally, $\varepsilon_m=0$ on the two~semi-infinite non-disordered chains defined by the sites $m>N$ and $m<1$, respectively.

Let $R_N$ and $T_N$ denote the reflection and transmission amplitudes for an electron wave incident from the right with wavenumber $-k$, and hence energy

\noindent
\begin{equation}\label{eq3}
E=2\cos k\;,\;o\leq k\leq \pi\quad .
\end{equation}

\noindent
Thus, by definition,

\begin{equation}\label{eq4}
\varphi_n= e^{-ik(n-N)}+R_N e^{ik(n-N)},\;\text{for}\; n\geq N\quad ,
\end{equation}

\noindent
and

\begin{equation}\label{eq5}
\varphi_n= T_N e^{-ikn},\;\text{for}\; n\leq 1\quad .
\end{equation}

\noindent
We then define the so-called Riccati ratios

\begin{equation}\label{eq6}
Y_n={\varphi_{n+1}\over\varphi_n}\quad ,
\end{equation}

\noindent
which reduce the Schr\"{o}dinger equation (\ref{eq2}) to the two-point recursion relation

\begin{equation}\label{eq7}
Y_n=E-\varepsilon_n-{1\over Y_{n-1}}\quad ,
\end{equation}

\noindent
where, from (\ref{eq4}), the variable $Y_N$ at the last site $N$ of the disordered section is related to $R_N$ by

\begin{equation}\label{eq8}
Y_N={e^{-ik}+R_Ne^{ik}\over1+R_N}\quad .
\end{equation}

\noindent
By comparing now the equations (\ref{eq4}-\ref{eq7}) for the chain with $N$ disordered sites with the corresponding equations for a chain with $N-1$ sites obtained
by removing the last disordered site ($\varepsilon_N=0$), and noting that in both cases $Y_0=e^{-ik}$, it follows that both chains have the same $Y_1,Y_2\ldots
Y_{N-1}$ but different $Y_N$.  This immediately leads to a recursion relation between $R_N$ and $R_{N-1}$, which is given by equation (\ref{eq7}) for $n=N$
where for $Y_N$ we insert (\ref{eq8}), and for $Y_{N-1}$ the analog of (\ref{eq8}) for the chain of $N-1$ sites since $Y_{N-1}$ is the same for the chain with $N$
sites and for the chain with $N-1$ sites.

\begin{equation}\label{eq9}
R_N={e^{2ik}R_{N-1}+i\nu_N(1+e^{2ik}R_{N-1})\over 1-i\nu_N(1+e^{2ik}R_{N-1})}\quad ,
\end{equation}

\noindent
where

\begin{equation}\label{eq10}
\nu_N={\varepsilon_N\over 2\sin k}\quad .
\end{equation}

We now extend the discussion of Barnes and Luck\cite{3} to obtain an analogous recursion relation for the transmission amplitude $T_N$.  This relation is found by
expressing $T_N=e^{ik}\varphi_1$, for the chain of length $N$ in terms of the Riccati ratios $Y_n$ using the identity
\linebreak $\varphi_1=\prod^{N-1}_{n=1}{1\over Y_n}(1+R_N)$, and $T_{N-1}=e^{ik}\varphi_1$, for the chain of length $N-1$ in terms of a similar identity.  Then,
by using the fact that the Riccati variables $Y_1,
\ldots,Y_{N-1}$ are the same for the chains of lengths $N$ and $N-1$ and taking the ratio $T_N/T_{N-1}$ we finally get

\begin{equation}\label{eq11}
T_N={(1+R_N)T_{N-1}\over e^{-ik}+e^{ik}R_{N-1}}\quad .
\end{equation}

\noindent
By using (\ref{eq9}), this expression is conveniently rewritten in the form

\newcounter{saveeqn}
\newcommand{\alpheqn}{\setcounter{saveeqn}{\value{equation}}%
\stepcounter{saveeqn}\setcounter{equation}{0}%
\renewcommand{\theequation}
{\mbox{\arabic{saveeqn}-\alph{equation}}}}
\newcommand{\reseteqn}{\setcounter{equation}{\value{saveeqn}}%
\renewcommand{\theequation}{\arabic{equation}}}

\setcounter{equation}{10}
\alpheqn
\begin{equation}\label{eq11a}
T_N={e^{ik}T_{N-1}\over 1-i\nu_N(1+e^{2ik}R_{N-1})}\quad .
\end{equation}

\reseteqn

The equations (\ref{eq9}) and (\ref{eq11a}) describe the variation of the reflection and transmission coefficients of a disordered chain when an extra
disordered site is added to it.  The initial conditions for these recursions are obvioulsy $R_0=0$ and $T_0=1$.  The unitarity relation for the reflection and
transmission coefficients, 

\begin{equation}\label{eq12}
|R_N|^2+|T_N|^2=1\quad ,
\end{equation}

\noindent
may be explicitely checked for a small $N$ e.g. $N=1,2,\ldots$  For example, for $N=1$ one has, from (\ref{eq9}) and (\ref{eq11a}),

\begin{equation}\label{eq13}
R_1={i\nu_1\over 1-i\nu_1}\;,\;T_1={e^{ik}\over 1-i\delta_1}\quad ,
\end{equation}

\noindent
which clearly obey (\ref{eq12}) and illustrate the growth of the reflection coefficient and the corresponding decrease of the transmission coefficient of an
incident wave, due to scattering by a single random site potential.

In order to derive the invariant imbedding equations for reflection and transmission amplitudes from Eqs (\ref{eq9}-\ref{eq11}), we first take the continuum
limit.  To this end we reinstate the lattice constant $a$ and redefine the chain length $L=Na$, the end-site energy $\varepsilon (L)\equiv\varepsilon_N$ and the
related variable $\nu(L)\equiv\nu_N$, as well as the reflection and transmission amplitudes $R(L)\equiv R_N$ and $T(L)\equiv T_N$ and, finally, we define the
finite differences $R'(L)={1\over a}\left[R(L)-R(L-a)\right]$ and $T'(L)={1\over a}\left[T(L)-T(L-a)\right]$.  The reinstatement of $a$ leads furthermore to the
replacement of solutions $\varphi_n=e^{ikn}$ for a non-disordered chain by $\varphi_n\equiv\varphi (na)=e^{ikna}$ and hence to
the replacement of $k$ by $ka$ in (\ref{eq3}) and in (\ref{eq9}-\ref{eq11}).  In the continuum limit, $a\rightarrow 0$,
$R'(L)$ and $T'(L)$ reduce to the corresponding derivatives $dR(L)/dL$ and $dT(L)/dL$, so that e.g. $R(L-a)=R(L)-a\;dR(L)/dL$.  Now
since the invariant imbedding equations are first order differential equations it follows that we have to restrict to
contributions up to order $a$ in the expansion of (\ref{eq9}) and (\ref{eq11}), where we thus approximate $e^{ika}$ by
$1+ika$.  Using these results the continuum limits of (\ref{eq9}) and (\ref{eq11}) become 

\begin{eqnarray}\label{eq14}
\lefteqn{a{dR(L)\over dL}= }\nonumber \\
 & {2ikaR(L)+i\nu(L)(1+R(L))(1+R(L)+2ikaR(L))\over 1+i\nu(L)(1+R(L))}\quad ,
\end{eqnarray}

\noindent
and

\begin{equation}\label{eq15}
a{dT(L)\over dL}=\biggl[ika+i\nu (L)\left(1+(1+2ika)R(L)-a{dR(L)\over dL}\right)\biggr]\quad .
\end{equation}

\noindent
Next, since the invariant imbedding equations are linear in the random potential (proportional to the site potential
$\varepsilon_N\equiv \varepsilon (L)$ in the present case) we wish to expand the r.h.s. of (\ref{eq14}-\ref{eq15}) to linear
order in $\nu_L$.  For this we require

\begin{equation}\label{eq16}
|\varepsilon(L)|<<2|\sin ka|\quad ,
\end{equation}

\noindent
or, within the approximation of (\ref{eq14}-\ref{eq15}), $|\varepsilon (L)|<<2ka$.  Clearly, this condition is not obeyed near
the band edges.  By expanding the r.h.s. of (\ref{eq14}-\ref{eq15}) to linear order in $\varepsilon (L)$ these equations may
finally be reduced to the form

\begin{equation}\label{eq17}
ik{dR(L)\over dL}=-2k^2R(L)-{\varepsilon(L)\over 2a^2}(1+R(L))^2\quad ,
\end{equation}

\noindent
and

\begin{equation}\label{eq18}
ik{dT(L)\over dL}=-k^2T(L)-{\varepsilon(L)\over 2a^2}(1+R(L))T(L)\quad ,
\end{equation}

\noindent
which coincide with the invariant imbedding stochastic equations\cite{17,18} for the amplitude reflection and transmission
coefficients for a 1D chain with a random potential

\begin{equation}\label{eq19}
V(L)=-{\varepsilon(L)\over 2a^2}\quad .
\end{equation}

We emphasize that the validity of Eqs (\ref{eq17}-\ref{eq18}) is subjected to the weak disorder condition (\ref{eq16}) for
wavenumbers $k$ different from the band edge values.  On the other hand, we recall that previous derivations of these
equations\cite{17,18} view the disordered chain as an inhomogeneous continuum and assume that the system of length $L$ is
imbedded invariantly in its extension of length $L+\Delta L$\cite{17}.  Also, these derivations do not involve an explicit
restriction to a weak random potential.

\subsection{Analysis of reflection phase distributions in 1D}\label{prtb}

A variety of detailed numerical and/or analytical results for distributions of reflection phases for weakly disordered
Anderson- or invariant imbedding models have been discussed in the literature\cite{2,4,8,10,12}.  Here we wish to present a
more complete comparison of these results in the perspective of the relation between these models established in \ref{prta}.  Our analysis shows that the
results for phase-distributions for chains described by the Anderson- and invariant imbedding models, respectively, are in good
qualitative agreement, as expected.

For the purpose of our discussion we refer to a series of figures in the literature where phase distributions for the above
models have been plotted over a phase interval of $2\pi$: we find it useful to classify these figures in tables~\ref{tabl1}(a)
and \ref{tabl1}(b) according to whether they correspond to low- or strong reflection systems, respectively.  A weakly
reflecting chain of length $L$ (which for the Anderson model is identified with the number of atomic sites, $N$) has a
localization length $L_c$ larger than $L$ while a strongly reflecting one has a localization length smaller than $L$. 
Our assignment of the numerical results of Refs~\cite{2,4,12} to the $L<<L_c$ or $L>>L_c$ categories is based on estimates of
the localization length using the familiar energy dependent expression for weak disorder, as given in these references.  We
recall that the curves of fig.~1 of Ref.~\cite{8} for low reflection systems and those of fig.~1 of Ref.~\cite{10} for strong
reflection systems are based on explicit analytic expressions for the phase distribution while the other figures referred to
in the tables are obtained from numerical calculations\cite{2,4,12}.  

In the low reflection case the uniform phase distributions in figs~2(b) and 2(d), for large $kL$, of Ref.~\cite{2} for the Anderson model are in good agreement
with corresponding nearly uniform distributions of fig.~1 of Ref.~\cite{8} for $kL=10^2$ and $kL=10^3$, for the invariant
imbedding model.  Similarly, the figs~2(a) and 2(c) of Ref.~\cite{2} and the fig.~1(b) of Ref.~\cite{4} for the Anderson
model, for lower $kL$ values, agree well with the analytic results of fig.~1 of Ref.~\cite{8} for the invariant imbedding
model, for $kL=5$ and for $kL=10$, which indicate a symmetric double peak phase distribution in intervals of $2\pi$.  On the
other hand, in the case of strong reflection (for weak disorder) the nearly uniform distributions in the fig.~1(b) of
Ref.~\cite{2} and in the fig.~1(a) of Ref.~\cite{4} for the Anderson model, for large $kL_c$ are in good agreement with the
distributions of fig.~1 of Ref.~\cite{10} for $kL_c=100$ and $kL_c=200$ and with the distribution of fig.~6(c) of
Ref.~\cite{12} for $kL_c=100$ and $kL_c=1000$, for the invariant imbedding model.  In the limit $kL_c<<1$ the general
expression for the stationary phase distribution obtained in\cite{9} has to be evaluated numerically.  In this limit both the
results of fig.~1(d) of Ref.~\cite{4} for the Anderson model and of fig.~6(a) of Ref.~\cite{12} for the invariant imbedding
model indicate that the phase distribution $P_\theta(\theta)$ is a symmetric function which is strongly peaked about
$\theta=\pi$.  Furthermore, as $kL_c$ is increased from zero the peak at $\theta=\pi$ shifts progressively to larger $\theta$
and a small secondary peak develops at $\theta<\pi$, so that the overall shape of the phase distribution becomes broader,
before tending, for $kL_c>>1$, to the results discussed above.  We note, in particular, the similarity between the results of
fig.~1 of Ref.~\cite{10} for moderately large $kL_c$ ($kL_c=10$ and $kL_c=20$) and the fig.~6(c) of Ref.~\cite{12} for
$kL_c=10$.  

In conclusion, the above discussion reveals excellent overall agreement between reflection phase distributions obtained,
respectively, form the Anderson model for weak disorder and from the invariant imbedding model.  This is expected to some
extent from the equivalence of these two~models in the continuum limit demonstrated in \ref{prta}.  We find, in
particular, that the discreteness of the Anderson tight-binding lattice does not significantly influence the qualitative form
of the phase distribution for weak disorder.

\subsection{Reflection coefficient in the Anderson model}\label{prtc}
In order to illustrate more precisely the effect of the discreteness of the Anderson model for weak disorder in a simple
typical case, we calculate the first and second moments, $\langle|R_N|^2\rangle$ and $\langle|R_N|^4\rangle$ of the
reflection coefficient.  These moments have previously been discussed within the invariant imbedding model\cite{8}.  In order
to find moments of $|R_N|^2$ for weak disorder we first obtain the exact solution for $R_N$ to linear order in the reduced
site energies (\ref{eq10}) by iterating (\ref{eq9}) with the initial condition $R_0=0$.  This yields

\begin{equation}\label{eq20}
R_N={i\over 2\sin k}\sum^{N}_{m=1}e^{2ik(N-m)}\varepsilon_m\quad .
\end{equation}

\noindent
Assuming the $\varepsilon_m$ to be independent gaussian variables with mean zero and correlation

\begin{equation}\label{eq21}
\langle\varepsilon_m\varepsilon_n\rangle=\varepsilon^2_0\delta_{m,n}\quad ,
\end{equation}

\noindent
we obtain successively from (\ref{eq20}):

\begin{equation}\label{eq22}
\langle|R_N|^2\rangle={\varepsilon_0^2N\over 4\sin^2 k}\quad ,
\end{equation}

\noindent
and

\begin{equation}\label{eq23}
\langle|R_N|^4\rangle={\varepsilon_0^4\over 16\sin^4 k}\left[2N^2+\left({\sin 2kN\over \sin 2k}\right)^2\right]\quad .
\end{equation}

In obtaining (\ref{eq23}) we have expressed averages of products of four~random energies in terms of averages of the form
(\ref{eq21}) using a well-known property of mutually independent gaussian variables\cite{19}. The corresponding results for a rectangular 
distribution of site energies between values
$-W$ and $W$ are obtained by replacing $\varepsilon^2_0$ by $W^2/3$.

The Eq.~(\ref{eq22}) is identified as usual with Ohm's law for the Landauer resistance for weak disorder,
$\rho\simeq\langle|R_N|^2\rangle=2N/L_c$, whose expression in terms of the localization length follows from (\ref{eq12}) and the definition $|T_N|\sim
e^{-N/L_c}$, in the limit $N<<L_c$.  From (\ref{eq22}) we have

\begin{equation}\label{eq24}
L_c={8\sin^2 k\over\varepsilon^2_0}\;,\;E=2\cos k\quad ,
\end{equation}

\noindent
which coincides with the well-known expression obtained by Thouless\cite{20}.  Its continuum limit,
$L_c=8k^2/\varepsilon^2_0$ reduces to the localization length $L_c=2k^2/V_0^2$ in a continuous gaussian potential with correlation parameter
$V_0=-\varepsilon_0/2$ (equation~\ref{eq19}).  In fact, in the continuum limit $(ka\rightarrow 0)$ the Eqs~(\ref{eq22}-\ref{eq23}) coincide with the results
obtained previously in an invariant- imbedding treatment\cite{8}.  As noted above the continuum limit breaks down near the band edges, $ka=0,\pi$. 
Finally, we recall that the second term on the r.h.s. of (\ref{eq23}) describes the effect of the inhomogeneity of the
reflection phase distribution\cite{8}.

\section{PHASE AND DELAY TIME RANDOMNESS FOR STRONG DISORDER}\label{sec3}

In this Section we discuss a simple limiting case of strongly disordered site energies in the Anderson model.  We consider a
random tight-binding chain where at every site the potential takes large values close to a fixed magnitude, which are
positive or negative at random.  This model corresponds to an extreme case of strong disorder since small random site energies
are a priori excluded.  Thus we assume the independently distributed site energies $\varepsilon_n$ to be described by
identical doubly peaked distributions with peaks centered at $\varepsilon_n=\pm c,c>>1$, with small standard deviation,
$\sigma<<1$.  Typical cases of such continuous distributions are the doubly-peaked gaussian

\begin{equation}\label{eq25}
p_\varepsilon(\varepsilon_n)={1\over 2\sigma\sqrt{2\pi}}
\left[e^{-{(\varepsilon_n-c)^2\over 2\sigma^2}}+e^{-{(\varepsilon_n+c)^2\over 2\sigma^2}}\right]\quad ,
\end{equation}

\noindent
and the doubly peaked rectangular distribution (with flat peaks centered at $\varepsilon_n=\pm c$ and standard deviation
${W\over \sqrt 3}<<1$)

\begin{eqnarray}\label{eq26}
p_\varepsilon(\varepsilon_n)&=&{1\over 4W}\left[\theta(\varepsilon_n-c+W)-\theta(\varepsilon_n-c-W)\right.\nonumber\\\
&+& \left. \theta (\varepsilon_n+c+W)-\theta(\varepsilon_n+c-W)\right]\quad .
\end{eqnarray}

\noindent
Note that the variances, $\text{var}\varepsilon_n=\sigma^2+c^2$, $\text{var}\varepsilon_n=W^2/3+c^2,c>>1$, of (\ref{eq25}) and
(\ref{eq26}), respectively, characterize the limit of strong disorder in the present case as do large variances in the corresponding single-peak
distributions (obtained by letting $c=0$), $\text{var}\varepsilon_n=\sigma^2$, with $\sigma>>1$, and
$\text{var}\varepsilon_n=W^2/3$, with $W/\sqrt 3>>1$, in the usual Anderson model.  The advantage of the above model of strong
disorder is that, due to the fact that typical values of the site energies $\varepsilon_n$ are large, we may use perturbation
theory in $1/\varepsilon_n$ for studying effects of the disorder.

We now analyse the reflection phase- and delay time distributions for the strongly disordered Anderson model with the
distributions (\ref{eq25}-\ref{eq26}) of site energies.  We shall discuss our analytical expressions in relation to recent
numerical results for these distributions in the ordinary Anderson model for strong disorder\cite{4}.  The probability
distribution of the phase of the reflection amplitude, $R_n=|R_n|e^{i\theta_n}$ and the corresponding distribution of the
Wigner delay time, $\tau_n=d\theta_N/dE$, will be found from the explicit solution of the stochastic recursion
relation (\ref{eq9}).  For strong disorder ($c>>1,\sigma<<1\;\text{or}\; W/\sqrt 3<<1$) we obtain by successive iteration of
(\ref{eq9})

\begin{equation}\label{eq27}
R_N =-1-{1\over i\nu_N}+{1\over\nu^2_N}{1\over1+e^{2ik}R_{N-1}}+\theta (\nu^{-3}_N)\quad ,
\end{equation}

\begin{equation}\label{eq28}
R_N =-1-{1\over i\nu_N}-{e^{-ik}\over 2i\nu^2_N \sin k}+\theta (\nu^{-3}_N,\nu^{-2}_N\nu^{-1}_{N-1})\quad .
\end{equation}

\noindent
Note that, to second order in $1/\nu_n$, $R_N$ has unit modulus, which implies that there is only backscattering to this
order.  Furthermore, to order $1/\nu^2_N$, $R_N$ is independent of scattering by interior sites $n<N$ and hence independent
of the length of the disordered chain.  In other words, the second order expression (\ref{eq28}) describes reflection in an
asymptotic stationary regime, $L>>L_c$.

For simplicity, we restrict to the linear order approximation

\setcounter{equation}{27}
\alpheqn
\begin{equation}\label{eq28a}
\quad R_N\simeq -1-{1\over i\nu_N}\quad ,
\end{equation}

\reseteqn

\noindent
for studying the phase- and delay time distribution, $P_\theta (\theta_N)$ and $P_\tau (\tau_N)$, respectively.  From
(\ref{eq28a}), (\ref{eq10}) and (\ref{eq3}) we obtain

\begin{equation}\label{eq29}
\theta_N\simeq -{2\sin k\over \varepsilon_N}\quad ,
\end{equation}
and
\begin{equation}\label{eq30}
\tau_N\simeq {\cot k\over \varepsilon_N}\quad .
\end{equation}

\noindent
where the distribution of $z=1/\varepsilon_N$ is given in terms of the site energy distributions (\ref{eq25}) or (\ref{eq26}), by

\begin{equation}\label{eq31}
p_z(z)={1\over z^2}\;p_\varepsilon \left({1\over z}\right)\quad .
\end{equation}

\noindent
Finally, from Eqs (\ref{eq29}-\ref{eq31}) we readily obtain the phase- and delay time distributions in the form

\begin{equation}\label{eq32}
P_\theta(\theta_N)={2|\sin k|\over \theta^2_N}p_\varepsilon\left(-{2\sin k\over \theta_N}\right)\quad ,
\end{equation}

\noindent
and

\begin{equation}\label{eq33}
P_\tau (\tau_N)={|\cot k|\over \tau^2_N}p_\varepsilon \left({\cot k\over \tau_N}\right)\quad .
\end{equation}

\noindent
From Eqs (\ref{eq32}) and (\ref{eq25}-\ref{eq26}) it follows that only small phases in the neighbourhood of the peak
positions, $\theta^\pm_N\simeq \pm{2|\sin k|\over c}$ have appreciable probability in our model of extreme strong disorder. 
We note that the Anderson model with an ordinary wide single-peak distribution of site energies also leads to a
characteristic double peak structure for the phase distribution (see the numerical results in fig.~1(c) of Ref.~\cite{4}). 
However, in contrast to (\ref{eq32}), the peak positions in the latter case are close to $\pi$ and their separation is of
order unity\cite{4}.

Turning to the delay time distribution (\ref{eq33}) we observe that delay time probabilities are small except near the
short-time maximum $\tau_N\sim{|\cot k|\over c}$.  In particular, the absence of an asymptotic power law tail,
$PÐ\tau\sim1/\tau^2$, as in (\ref{eq1}), is due to the fact that in the present case an incoming particle (wave) does not
penetrate into the disordered sample beyond the outermost site at which it is reflected (equation (\ref{eq28a})).

Ossipov {\it et al.}\cite{4} have also studied numerically the delay time distribution for the Anderson model with strong
disorder i.e. for a wide single-peak distribution of site energies.  They found a long time tail
$P_\tau(\tau)\sim1/\tau^2,\;\tau\rightarrow\infty$, which is believed to be associated with Azbel resonances.  From this
they concluded that the long-time tail exists regardless of the strength of disorder.  This conclusion is partly disproved,
however, by our study of the above model of extreme strong disorder.  Indeed we find that the long-time probability is
strongly reduced by an exponentially small factor, $\exp (-c^2/2\sigma^2)<<1$, due to the fact that the incoming particle is
reflected right at the edge of the disordered sample.

\end{multicols}

\begin{table}
\caption{Classification of literature results for $P_\theta(\theta)$ pertaining (a) to the low reflection regime, $L(N)\ll L_c$, and (b) to the
strong reflection regime, $L(N)\gg L_c$, in the invariant imbedding model and in the Anderson model for weak disorder.\label{tabl1}}
\begin{tabular}{l|c|l|c}
\multicolumn{4}{l}{(a)\qquad\qquad\qquad $L(N)<<L_c$}\\
\tableline
Anderson Model & Reference & Invariant imbedding & Reference\\
\tableline
fig.~2(a), $N=10,\;k=0,46\pi,\;L_c\simeq 6.4\;10^6$ & 2 & fig.~1, $kL=5$ & 8\\
fig.~2(b), $N=50,\;k=0,46\pi,\;L_c\simeq 2.4\;10^7$ & 2 & fig.~1, $kL=10$ & 8\\
fig.~2(c), $N=10,\;k=0,46\pi,\;L_c\simeq 2.4\;10^3$ & 2 & fig.~1, $kL=100$ & 8\\
fig.~2(d), $N=50,\;k=0,46\pi,\;L_c\simeq 6.4\;10^3$ & 2 & fig.~1, $kL=1000$ & 8\\
fig.~1(b), $k=\pi/2,\;L_c\simeq 2.4\;10^3$          & 4 &                             & 8\\
\tableline
\multicolumn{4}{l}{(b)\qquad\qquad\qquad $L(N)>>L_c$}\\
\tableline
Anderson Model & Reference & Invariant imbedding & Reference\\
\tableline
fig.~1(a), $N=10^3,\;k=2\pi/3,\;L_c=290$   & 2 & fig.~1,    $kL_c=10$ & 10\\
fig.~1(b), $N=10^4,\;k=\pi/2,\;L_c=390$    & 2 & fig.~1,    $kL_c=20$ & 10\\
fig.~1(a), $k=\sqrt\pi,\;L_c\simeq 60$     & 4 & fig.~1,    $kL_c=100$ & 10\\
fig.~1(d), $k=0.001\sqrt\pi,\;L_c<10^{-6}$ & 4 & fig.~1,    $kL_c=200$ & 10\\
 &  																																																				& fig.~6(a), $kL_c=0.00001$ & 12\\
 &  																																																				& fig.~6(a), $kL_c=0.0001$ & 12\\
 &  																																																				& fig.~6(b), $kL_c=0.01$ & 12\\
 &  																																																				& fig.~6(c), $kL_c=10$ & 12\\
 &  																																																				& fig.~6(c), $kL_c=100$ & 12\\
 &  																																																				& fig.~6 (c), $kL_c=1000$ & 12
\end{tabular}
\end{table}


\newpage
\begin{thebibliography}{99}
\bibliographystyle{unsrt}
\bibitem{1} C.J.~Lambert and M.F.~Thorpe, Phys.~Rev.~B{\bf 26}, 4742 (1982).
\bibitem{2} A.D.~Stone, D.C.~Allan, and J.D.~Joanopoulos, Phys.~Rev.~B{\bf 27}, 836 (1983).
\bibitem{3} C.~Barnes and J.M.~Luck, J.~Phys. A{\bf 23}, 1717 (1990).
\bibitem{4} A.~Ossipov, T.~Kotos, and T.~Geisel, Phys.~Rev.~B{\bf 61}, 11411 (2000); F.~Steinbach, A.~Ossipov, T.~Kotos, and T.~Geisel, Phys.~Rev.~Lett. {\bf 85},
4426 (2000).
\bibitem{5} See also the review by J.B.~Pendry, Adv.~Phys. {\bf 43}, 461 (1994).
\bibitem{6} S.K.~Joshi, A.K.~Gupta, and A.M.~Jayannavar, Phys.~Rev.~B{\bf 58}, 1092 (1998).
\bibitem{7} P.L.~Sulem, Physica {\bf 70}, 190 (1973).
\bibitem{8} J.~Heinrichs, J.~Phys.~C: Solid~State~Phys. {\bf 21}, L153 (1988).
\bibitem{9} J.~Heinrichs, J.~Phys.: Condensed~Matter {\bf 2}, 1559 (1990).
\bibitem{10} J.~Heinrichs, Solid~State~Comm. {\bf 76}, 543 (1990). 
\bibitem{11} A.M.~Jayannavar, Solid~State~Comm. {\bf 73}, 247 (1990).
\bibitem{12} Kichong~Kim, Phys.~Rev.~B{\bf 58}, 6153 (1998).
\bibitem{13} A.M.~Jayannavar, G.V.~Vijayagovindan, and N.~Kumar,Z.~Phys.~B{\bf 75}, 77 (1989). 
\bibitem{14} A.~Comtet and C.~Texier, J.~Phys.~A{\bf 30}, 8017 (1997)(a); J.~Texier and A.~Comtet, Phys.~Rev.~Lett., {\bf 82}, 4220
(1999)(b). 
\bibitem{15} S.~Anantha~Ramakrishna and N.~Kumar, Phys.~Rev.~B{\bf 61}, 3163 (2000).
\bibitem{16} Y.V.~Fyodorov and H.J.~Sommers, J.~Math.~Phys. {\bf 38}, 1918 (1997).
\bibitem{17} R.~Bellman and G.M.~Wing, An Introduction to Invariant Imbedding (Wiley, New~York, 1976).
\bibitem{18} For an extensive review of invariant imbedding and of applications, see R.~Rammal and B.~Doucot, J.~Phys. (Paris) {\bf 48}, 509 (1987);
B.~Doucot and R.~Rammal, ibid. {\bf 48}, 527 (1987).
\bibitem{19} See e.g., N.G.~Van~Kampen, Stochastic Processes in Physics and Chemistry (North-Holland, Amsterdam, 1981).
\bibitem{20} D.J.~Thouless, in Ill-Condensed Matter, edited by R.~Balian, R.~Maynard and G.~Toulouse (North-Holland, Amsterdam, 1979).
\end{thebibliography}
\end{document}